\documentclass[aps,prd,twocolumn,superscriptaddress,showpacs]{revtex4}

\RequirePackage{lineno}

\usepackage{epsfig}
\usepackage{graphicx}
\usepackage{dcolumn}
\usepackage{bm}
\usepackage{setspace}

\begin{document}




\title{Observation of the Decay $B_c^{\pm} \rightarrow J/\psi \, \pi^{\pm}$
and Measurement of the $B_c^{\pm}$ Mass}
\affiliation{Institute of Physics, Academia Sinica, Taipei, Taiwan 11529, Republic of China} 
\affiliation{Argonne National Laboratory, Argonne, Illinois 60439} 
\affiliation{Institut de Fisica d'Altes Energies, Universitat Autonoma de Barcelona, E-08193, Bellaterra (Barcelona), Spain} 
\affiliation{Baylor University, Waco, Texas  76798} 
\affiliation{Istituto Nazionale di Fisica Nucleare, University of Bologna, I-40127 Bologna, Italy} 
\affiliation{Brandeis University, Waltham, Massachusetts 02254} 
\affiliation{University of California, Davis, Davis, California  95616} 
\affiliation{University of California, Los Angeles, Los Angeles, California  90024} 
\affiliation{University of California, San Diego, La Jolla, California  92093} 
\affiliation{University of California, Santa Barbara, Santa Barbara, California 93106} 
\affiliation{Instituto de Fisica de Cantabria, CSIC-University of Cantabria, 39005 Santander, Spain} 
\affiliation{Carnegie Mellon University, Pittsburgh, PA  15213} 
\affiliation{Enrico Fermi Institute, University of Chicago, Chicago, Illinois 60637} 
\affiliation{Comenius University, 842 48 Bratislava, Slovakia; Institute of Experimental Physics, 040 01 Kosice, Slovakia} 
\affiliation{Joint Institute for Nuclear Research, RU-141980 Dubna, Russia} 
\affiliation{Duke University, Durham, North Carolina  27708} 
\affiliation{Fermi National Accelerator Laboratory, Batavia, Illinois 60510} 
\affiliation{University of Florida, Gainesville, Florida  32611} 
\affiliation{Laboratori Nazionali di Frascati, Istituto Nazionale di Fisica Nucleare, I-00044 Frascati, Italy} 
\affiliation{University of Geneva, CH-1211 Geneva 4, Switzerland} 
\affiliation{Glasgow University, Glasgow G12 8QQ, United Kingdom} 
\affiliation{Harvard University, Cambridge, Massachusetts 02138} 
\affiliation{Division of High Energy Physics, Department of Physics, University of Helsinki and Helsinki Institute of Physics, FIN-00014, Helsinki, Finland} 
\affiliation{University of Illinois, Urbana, Illinois 61801} 
\affiliation{The Johns Hopkins University, Baltimore, Maryland 21218} 
\affiliation{Institut f\"{u}r Experimentelle Kernphysik, Universit\"{a}t Karlsruhe, 76128 Karlsruhe, Germany} 
\affiliation{Center for High Energy Physics: Kyungpook National University, Daegu 702-701, Korea; Seoul National University, Seoul 151-742, Korea; Sungkyunkwan University, Suwon 440-746, Korea; Korea Institute of Science and Technology Information, Daejeon, 305-806, Korea; Chonnam National University, Gwangju, 500-757, Korea} 
\affiliation{Ernest Orlando Lawrence Berkeley National Laboratory, Berkeley, California 94720} 
\affiliation{University of Liverpool, Liverpool L69 7ZE, United Kingdom} 
\affiliation{University College London, London WC1E 6BT, United Kingdom} 
\affiliation{Centro de Investigaciones Energeticas Medioambientales y Tecnologicas, E-28040 Madrid, Spain} 
\affiliation{Massachusetts Institute of Technology, Cambridge, Massachusetts  02139} 
\affiliation{Institute of Particle Physics: McGill University, Montr\'{e}al, Canada H3A~2T8; and University of Toronto, Toronto, Canada M5S~1A7} 
\affiliation{University of Michigan, Ann Arbor, Michigan 48109} 
\affiliation{Michigan State University, East Lansing, Michigan  48824} 
\affiliation{University of New Mexico, Albuquerque, New Mexico 87131} 
\affiliation{Northwestern University, Evanston, Illinois  60208} 
\affiliation{The Ohio State University, Columbus, Ohio  43210} 
\affiliation{Okayama University, Okayama 700-8530, Japan} 
\affiliation{Osaka City University, Osaka 588, Japan} 
\affiliation{University of Oxford, Oxford OX1 3RH, United Kingdom} 
\affiliation{University of Padova, Istituto Nazionale di Fisica Nucleare, Sezione di Padova-Trento, I-35131 Padova, Italy} 
\affiliation{LPNHE, Universite Pierre et Marie Curie/IN2P3-CNRS, UMR7585, Paris, F-75252 France} 
\affiliation{University of Pennsylvania, Philadelphia, Pennsylvania 19104} 
\affiliation{Istituto Nazionale di Fisica Nucleare Pisa, Universities of Pisa, Siena and Scuola Normale Superiore, I-56127 Pisa, Italy} 
\affiliation{University of Pittsburgh, Pittsburgh, Pennsylvania 15260} 
\affiliation{Purdue University, West Lafayette, Indiana 47907} 
\affiliation{University of Rochester, Rochester, New York 14627} 
\affiliation{The Rockefeller University, New York, New York 10021} 
\affiliation{Istituto Nazionale di Fisica Nucleare, Sezione di Roma 1, University of Rome ``La Sapienza," I-00185 Roma, Italy} 
\affiliation{Rutgers University, Piscataway, New Jersey 08855} 
\affiliation{Texas A\&M University, College Station, Texas 77843} 
\affiliation{Istituto Nazionale di Fisica Nucleare, University of Trieste/\ Udine, Italy} 
\affiliation{University of Tsukuba, Tsukuba, Ibaraki 305, Japan} 
\affiliation{Tufts University, Medford, Massachusetts 02155} 
\affiliation{Waseda University, Tokyo 169, Japan} 
\affiliation{Wayne State University, Detroit, Michigan  48201} 
\affiliation{University of Wisconsin, Madison, Wisconsin 53706} 
\affiliation{Yale University, New Haven, Connecticut 06520} 
\author{T.~Aaltonen}
\affiliation{Division of High Energy Physics, Department of Physics, University of Helsinki and Helsinki Institute of Physics, FIN-00014, Helsinki, Finland}
\author{J.~Adelman}
\affiliation{Enrico Fermi Institute, University of Chicago, Chicago, Illinois 60637}
\author{T.~Akimoto}
\affiliation{University of Tsukuba, Tsukuba, Ibaraki 305, Japan}
\author{M.G.~Albrow}
\affiliation{Fermi National Accelerator Laboratory, Batavia, Illinois 60510}
\author{B.~\'{A}lvarez~Gonz\'{a}lez}
\affiliation{Instituto de Fisica de Cantabria, CSIC-University of Cantabria, 39005 Santander, Spain}
\author{S.~Amerio}
\affiliation{University of Padova, Istituto Nazionale di Fisica Nucleare, Sezione di Padova-Trento, I-35131 Padova, Italy}
\author{D.~Amidei}
\affiliation{University of Michigan, Ann Arbor, Michigan 48109}
\author{A.~Anastassov}
\affiliation{Rutgers University, Piscataway, New Jersey 08855}
\author{A.~Annovi}
\affiliation{Laboratori Nazionali di Frascati, Istituto Nazionale di Fisica Nucleare, I-00044 Frascati, Italy}
\author{J.~Antos}
\affiliation{Comenius University, 842 48 Bratislava, Slovakia; Institute of Experimental Physics, 040 01 Kosice, Slovakia}
% ...
\collaboration{CDF Collaboration\footnote{With visitors from $^a$University of Athens, 15784 Athens, Greece, 
$^b$Chinese Academy of Sciences, Beijing 100864, China, 
$^c$University of Bristol, Bristol BS8 1TL, United Kingdom, 
$^d$University Libre de Bruxelles, B-1050 Brussels, Belgium, 
$^e$University of California Irvine, Irvine, CA  92697, 
$^f$University of California Santa Cruz, Santa Cruz, CA  95064, 
$^g$Cornell University, Ithaca, NY  14853, 
$^h$University of Cyprus, Nicosia CY-1678, Cyprus, 
$^i$University College Dublin, Dublin 4, Ireland, 
$^j$University of Edinburgh, Edinburgh EH9 3JZ, United Kingdom, 
$^k$University of Heidelberg, D-69120 Heidelberg, Germany, 
$^l$Universidad Iberoamericana, Mexico D.F., Mexico, 
$^m$University of Manchester, Manchester M13 9PL, England, 
$^n$Nagasaki Institute of Applied Science, Nagasaki, Japan, 
$^o$University de Oviedo, E-33007 Oviedo, Spain, 
$^p$Queen Mary, University of London, London, E1 4NS, England, 
$^q$Texas Tech University, Lubbock, TX  79409, 
$^r$IFIC(CSIC-Universitat de Valencia), 46071 Valencia, Spain, 
}}
\noaffiliation


\begin{abstract}

The $B_c^{\pm}$ meson
is observed through the decay
$B_c^{\pm} \rightarrow J/\psi \, \pi^{\pm}$, 
in data 
corresponding to an integrated luminosity of 2.4 fb$^{-1}$ recorded by
the CDF II detector at the Fermilab Tevatron.
A signal of $108\pm15$ candidates is observed, with a
significance that exceeds $8 \, \sigma$.
The mass of the $B_c^{\pm}$\ meson is measured to be
6275.6~$\pm$~2.9~(stat.)~$\pm$~2.5~(syst.)~MeV/$c^2$. 

\end{abstract}
\maketitle

The $B_c^{-}$\ meson \cite{conjugate} is composed of a bottom quark, $b$,
and an anti-charm quark, $\overline{c}$,
the heaviest quark flavors expected to  form mesons. 
The presence of two relatively heavy quarks is unique in 
the $B_c^{\pm}$\ system, and affects the theoretical calculation
of the decay 
properties and mass of the $B_c^{\pm}$\ meson. 
Either quark in the $B_c^{\pm}$\ meson can decay weakly,
which suggests copious decay 
modes \cite{theorydecays} and an expected lifetime much shorter than 
that of other
$B$ mesons \cite{theorylife}. The mass of the $B_c^{\pm}$\ meson has 
been predicted using a variety of theoretical techniques. Nonrelativistic 
potential models have been used to predict a mass of the $B_c^{\pm}$\ in the 
range of 6247 - 6286~MeV/$c^2$\ \cite{potentialbc},
and a slightly higher value is found for a perturbative QCD
calculation \cite{higherqcd}. Recent lattice QCD
calculations provide a $B_c^{\pm}$ mass prediction of 
$6304~\pm~12~^{+18}_{-0}$~MeV/$c^2$ \cite{latticebc}. 
Precision measurements of the properties of the $B_c^{\pm}$ are needed
to test these calculations.

The $B_c^{\pm}$\ meson is too massive to be produced at 
$e^+ \, e^-$ colliders 
operating near the $\Upsilon$(4S) center of mass energy, and only a few 
candidate events, which were consistent with background processes, were 
observed at LEP \cite{lep}. With a large $b$-quark production cross section 
and powerful multipurpose detectors, the Fermilab Tevatron is well-suited
to the study of all species of $B$ hadrons, including the $B_c^{\pm}$\ meson. 
In particular, this collaboration 
observed the $B_c^{\pm}$\ meson in semileptonic decays and
used these candidates to measure the lifetime
of the $B_c^{\pm}$ to be $0.463 ^{+0.073}_{-0.065} \pm 0.036$ ps 
\cite{bclife}. 
In addition, we have previously presented evidence for a 
$B_c^{\pm} \rightarrow J/\psi \, \pi^{\pm}$\ signal
\cite{saverioprl}. 

In this Letter, an analysis is presented 
which observes the decay mode $B_c^{\pm} \rightarrow J/\psi \, \pi^{\pm}$, with
$J/\psi \rightarrow \mu^+ \, \mu^-$, 
and precisely measures the $B_c^{\pm}$\ mass.
This result provides
the first observation of the $B_c^{\pm}$\ meson
in a fully reconstructed mode.
This observation is made in
$p\overline{p}$ collisions 
at a center of mass energy of 1.96 TeV using the
Collider Detector at Fermilab (CDF II) \cite{CDF_detector}.
The results presented here are based on 
a data sample with an integrated luminosity of 2.4 fb$^{-1}$,
and supercede our earlier measurement \cite{saverioprl}, 
which was based on an integrated luminosity of 360 pb$^{-1}$.  

This analysis makes use of the tracking and
muon identification systems.
The tracking system consists of 
double-sided silicon detectors \cite{CDF_SVXII} 
and a 96 layer open-cell drift 
chamber (COT) \cite{CDF_COT} 
that operate inside a solenoid with a 1.4 T field oriented
along the beam axis. 
Charged particles
originating from the collision point are measured in the tracking system
with a momentum resolution of 
$\sigma(p_T)/p_T \sim 0.0015 \, p_T$(GeV/$c)^{-1}$ \cite{pt_def}.
Muon candidates from the decay $J/\psi \rightarrow \mu^+ \, \mu^-$
are  identified by  two sets of drift  chambers located radially
outside  the   electromagnetic and   hadronic  calorimeters.
The central muon chambers
cover the pseudorapidity region $ |\eta| <  0.6$ 
and  are sensitive to muons with 
transverse momentum $p_T > 1.4 $ GeV/$c$.
A second muon system covers the region $  0.6 < |\eta| <  1.0$ and
detects  muons having $p_T > 2.0 $ GeV/$c$.  
Muon triggering and identification
are based on matching tracks measured in the muon system to COT tracks.
This analysis is based on events recorded with a trigger
that is dedicated to the collection of a 
$J/\psi \rightarrow \mu^+ \, \mu^-$ sample.
The first level of the three-level trigger system requires two
muon candidates with tracks in the COT and muon chamber systems
that match in the transverse view.
The second level imposes the requirement that muon candidates
have opposite charge 
and limits the accepted range of opening angle
between the two muon candidates.  The highest level 
of the $J/\psi$  trigger reconstructs the muon pair in software, and 
requires that the invariant mass of the pair falls within the range
$2.7-4.0$ GeV/$c^2$.

The analysis of the data begins with a selection of well-measured
$J/\psi \rightarrow \mu^+ \, \mu^-$ candidates.
Events are required to contain two
 oppositely  charged  muon candidates 
 that satisfy restrictive
 matching requirements, consistent with the full measurement
precision of the COT and muon chamber systems.
 We also require that both muon tracks have associated 
measurements in at least three layers of the silicon detector and
a two-track invariant  mass within 70 MeV/$c^2$ of the
world-average  $J/\psi$ mass \cite{PDG}. 
This data sample provides approximately 17 million events containing $J/\psi$
candidates, measured with an average mass resolution of 13 MeV/$c^2$.

Both $B^{\pm} \rightarrow J/\psi \, K^{\pm}$
and $B_c^{\pm} \rightarrow J/\psi \, \pi^{\pm}$ combinations are
reconstructed in this analysis.  The relatively plentiful
$B^{\pm}$ sample serves as a reference signal and is used
to develop criteria for the $B_c^{\pm}$ selection .
These final states are identified by assigning the $\pi^{\pm}$ or
$K^\pm$ mass to all tracks not used in the $J/\psi$ reconstruction.
In order to provide the measurement resolution sufficient to discriminate 
between directly produced tracks and $B$-hadron decay products,
the $\pi^{\pm}$ and $K^{\pm}$ candidate tracks are required to 
have measurements on at least three layers of the silicon detector.
Each three-track combination must satisfy 
a fit in which the tracks are required to originate from a common
vertex and the invariant mass of the muon pair is constrained to the
world average $J/\psi$ mass.
Approximately $65 \, 000 \: B^\pm$ candidates are identified in this 
loosely selected sample.

The selection requirements are further improved to give a large
$B^{\pm} \rightarrow J/\psi \, K^{\pm}$\ signal with very small 
background, in order to increase
the significance of the $B_c^{\pm}$\ observation. 
Since the $B_c^{\pm}$ lifetime is significantly shorter than the
lifetime of the $B^{\pm}$, each selection variable
under consideration has been studied for its effect on preserving signal
and removing background only for $J/\psi \, K^{\pm}$
combinations with 
$80 < ct < 300 \: \mu$m, where $t$ is the proper decay time determined from
$t \equiv \vec{L_T} \cdot \vec{p_T}(B) \frac{M(B)}{|p_T(B)|^2}$, 
$M(B)$ is the mass of the combination,
$\vec{p_T}(B)$ is the transverse momentum of the combination, and
$\vec{L_T}$ is the transverse displacement 
of the $J/\psi \, K^{\pm}$ decay vertex from the beam-line. 
Each selection quantity considered is evaluated for its efficiency in 
retaining the $B^{\pm}$ signal and reducing the combinatorial 
background in the $5.4 - 5.5$ GeV/$c^2$ mass range.

Several characteristics of the $J/\psi \, K^{\pm}$ candidates
are considered for selection requirements.
Minimum $p_T$ requirements 
on the $K^\pm$ and $B^\pm$  candidates are used to suppress
backgrounds from $J/\psi \, K^{\pm}$ combinations
that are not related to the $B^\pm$ decay.
Reasonable vertex quality is assured by placing a minimum value on the
accepted probability $P(\chi^2)$ of the mass- and vertex-constrained 
fit used to obtain the $B^\pm$ candidate.
The trajectory of the $K^\pm$  is required to originate from the
$B^\pm$ decay vertex by placing a requirement on its 
distance of closest approach
$d_{SV}(K)$ and associated uncertainty $\sigma_{d_{SV}}(K)$
with respect to the vertex found in the 
$J/\psi$ fit.
Similar quantities $d_{PV}(K)$ and $\sigma_{d_{PV}}(K)$
measured with respect to the primary vertex are used to remove tracks
that originate from direct production processes.
We suppress the promptly-produced combinatorial background by rejecting 
candidates with small $ct$. 
A requirement on  $\sigma_{ct}$
removes poorly-reconstructed combinations and other backgrounds.
We also reject $J/\psi \, K^{\pm}$ combinations that are 
inconsistent with having
originated from the beamline by requiring a small magnitude for
the transverse impact of the candidate,
$\vec{d}_{PV}(B) \equiv \vec{L_T} \times \vec{p_T}(B)/|p_T(B)|$, and a small
angle $\beta$ between  $\vec{L_T}$ and $\vec{p_T}(B)$.

\begin{figure}[p]
\psfig{figure=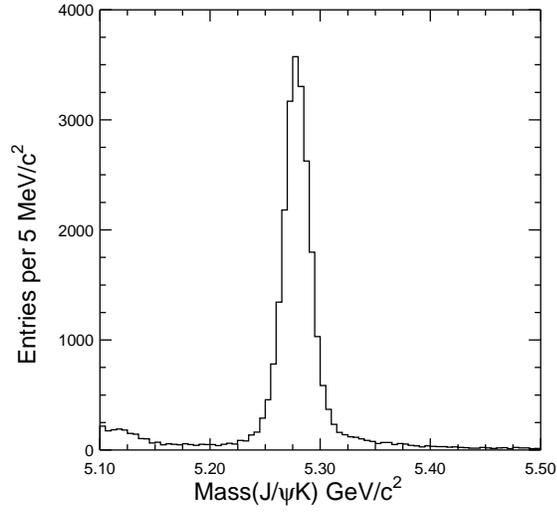, width=3.0in}
\caption{The invariant mass distribution of $J/\psi \, K^\pm$ combinations.
\label{fig:fig1} }
\end{figure}

Two sets of selection criteria based on these quantities are listed in 
Table \ref{table:selection}.  The first step of the selection process is to
impose the ``standard" selection
requirements listed above and retain all $J/\psi \, K^\pm$ 
combinations that satisfy them.
However, the selection studies indicate that a substantial number of 
$B^{\pm} \rightarrow J/\psi \, K^{\pm}$\ signal  events fail exactly
one of the standard selection requirements. 
To increase the overall signal efficiency,
we introduce a  ``high-$p_T$" selection, comprised of 
combinations that satisfy more restrictive $p_T$ requirements,
and pass all but one of
the other high-$p_T$ selection criteria in Table \ref{table:selection}. 
Figure~\ref{fig:fig1} shows the reconstructed
$B^{\pm} \rightarrow J/\psi \, K^{\pm}$\ signal for candidates 
that satisfy the  
standard or high-$p_T$ selection criteria. 
The additional combinations retained by the 
high-$p_T$\ selection increase the $B^\pm$\ yield by 28\%.
The total signal of approximately
$21 \, 100 \: B^\pm$\ candidates with a small background 
of 430 events in the $B^\pm$ sideband
region between 5.4 and 5.5~GeV/$c^2$\ demonstrates that this selection 
effectively removes the background to $B$-hadron candidates.

\begin{table}[ht]
\begin{center}
\caption{Selection variables and requirements for the standard selection
and high-$p_T$\ selection as
described in the text. Here ``$Trk$'' refers to the track combined with
the $J/\psi$\ and may be a $K^\pm$ or $\pi^{\pm}$ candidate for the 
$B^\pm$\ or $B_c^{\pm}$\ respectively. \label{table:selection}}
\begin{tabular}{rcc}
\hline \hline
Selection variable & Standard & High-$p_T$ \\
\hline
 $p_T(Trk)$ & $>$ 1.7~GeV/$c$ & $>$ 2.5~GeV/$c$ \\
 $p_T(J/\psi\ Trk)$ & $>$ 5~GeV/$c$ & $>$ 6~GeV/$c$ \\
\hline
 $P(\chi^{2})$ & $> 0.1\% $ & $> 1\% $ \\
 $|d_{SV}(Trk)|$  & $< 100~\mu$m & $< 80~\mu$m \\
 $|d_{PV}(Trk)|/ \sigma_{d_{PV}(Trk)}$ & $> 2.5 $ & $> 3$\\
 $|d_{PV}(B)|/ \sigma_{d_{PV}(B)}$  & $< 2.5$ & $< 2$ \\
 $ct$ & $> 80~\mu$m & $> 100~\mu$m \\
 $\sigma_{ct}$ & $< 30~\mu$m & $< 25~\mu$m \\
 $\beta$ & $< 0.4$ radians & $< 0.3$ radians \\
\hline \hline
\end{tabular}
\end{center}
\end{table}

\begin{figure}[p]
\psfig{figure=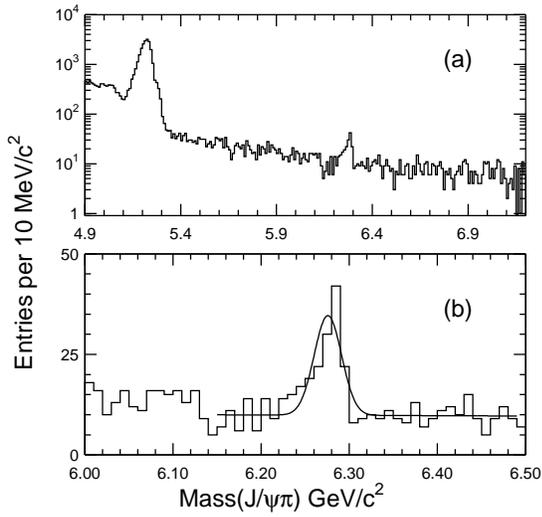, width=3.0in}
\caption{(a).  The invariant mass distribution of $J/\psi \, \pi^\pm$
combinations.  (b)  Identical to (a), but in
a narrower mass range around the theoretically favored $B_c^{\pm}$
mass.  The projection of the fit to the data is indicated by the 
curve overlaid on (b).
\label{fig:fig2} }
\end{figure}

The candidates used for the reconstruction of the $B_c^\pm$
are obtained by applying the selection criteria in 
Table \ref{table:selection}
to the sample interpreted as $J/\psi \, \pi^{\pm}$ .
Figure~\ref{fig:fig2} shows the  
$B_c^\pm \rightarrow J/\psi \, \pi^\pm$ candidates under the combined
standard and high-$p_T$\ selections. 
The data are also shown for a limited mass region suggested by 
theoretical expectations \cite{potentialbc,higherqcd,latticebc}.
A clear excess of $J/\psi \, \pi^{\pm}$ combinations
is evident around 6280~MeV/$c^2$.

The data shown in Fig. \ref{fig:fig2}(b) are fitted using an
unbinned log likelihood function 
that uses a Gaussian shape for the signal plus a 
background model that includes contributions from both combinatorial
sources and contributions from 
Cabibbo-suppressed $B_c^{\pm} \rightarrow J/\psi \, K^\pm$ decays
in the probability distribution function.
Simulated events are used to estimate the mass distribution expected from
the misidentified Cabibbo-suppressed process, 
and it is found to be approximately Gaussian, centered
60 MeV/$c^2$  below the true $B_c^{\pm}$ mass and having
a 30 MeV/$c^2$ width.
The characteristic width for the signal in the
probability distribution function  
is varied for each candidate in proportion to the 
mass uncertainty estimated from the individual track uncertainties.
A similar fit performed on the $B^\pm$ candidates finds that the
mass resolution is underestimated by a factor of $1.55$ with this method.
Consequently, the mass uncertainties used in the 
$J/\psi \, \pi^{\pm}$ fit are scaled by this factor.
For the central value of the fit, the Cabibbo-suppressed
contribution is fixed to 
0.05 of the total $B_c^\pm$ yield, which is comparable
to the measurements obtained for Cabibbo-suppressed $B^\pm$ decays \cite{PDG}.
Fig. \ref{fig:fig2}(b) shows a projection of
the fit overlaid on the data, which has $\chi^2/DoF = 34.6/32$
and a probability of $P(\chi^2)=0.30$.

The result of the unbinned fit 
over the mass range $6150 - 6500$ ~MeV/$c^2$
gives a $B_c^\pm$ signal of 
108~$\pm$~15 candidates with a mass of $6275.6~\pm~2.9$~MeV/$c^2$.
This fit was repeated over the same mass range
with the constraint that no signal is present
in order to obtain a significance measurement.
The ratio of likelihoods between the two fits
yields a probability for the null
hypothesis of $1.2\times 10^{-16}$, equivalent to a fluctuation of 
$8\sigma$ or more in a Gaussian distribution.
A second significance test was also performed, which
used a Monte Carlo simulation to 
estimate the probability for a background-only sample
to mimic a signal at least as significant as the one observed.
This study confirmed the significance estimate obtained 
from ratio-of-likelihoods test.  
We conclude that this enhancement
is an observation of the process $B_c^\pm \rightarrow J/\psi \, \pi^\pm$. 

Studies that were performed for the  mass
measurement of other $B$ mesons 
that contain a $J/\psi$ in the final state \cite{bmassrunii} are used
for an evaluation of some of the systematic uncertainties.
The systematic uncertainty due to tracking detector misalignments
and material distributions used in track fitting is assessed 
by varying these within reasonable limits.
Based on these studies, a systematic uncertainty of 
0.6 MeV/$c^2$ is assigned to account for misalignments and material modeling.
A comparison between the fitted $B^{\pm}$\ 
mass in 12 independent periods of data 
finds the measured mass of the $B^\pm$\ meson 
for each subset to be within 1.8$\sigma$ of the entire sample.
Consquently, no systematic uncertainty is assigned for 
time dependent effects on this mass measurement. 
The momentum scale of the tracking system is calibrated with the
$J/\psi$, $\psi(2S)$, and $\Upsilon$ states,
all of which have been well measured previously \cite{PDG}.  
The uncertainty of the 
momentum scale for the decay products of $B_c^{\pm}$  
which was derived from this calibration contributes
an additional mass uncertainty of 0.6~MeV/$c^2$. 
The sensitivity of the mass measurement to the assumed Cabibbo-suppressed
background process was tested with simulated events.  Variations of
$0.0-0.1$ for the fractional 
contribution correspond to mass measurement variations
of 0.8~MeV/$c^2$, which is taken to be the systematic uncertainty 
associated with this unknown background contribution.

An additional systematic uncertainty comes from the fitting
procedure. The scale factor associated with
the mass uncertainty which is used in the fit was set to the best value
obtained in a fit to the topologically similar
$B^{\pm} \rightarrow J/\psi \, K^{\pm}$ final state. 
However, other reconstructed $B$ hadrons
that include a $J/\psi$ in the final state
have been studied as well.
Scale factors in the range of 1.25 - 2.0 have been 
suggested by these studies, 
so the fit for the $B_c^{\pm}$ mass
was repeated for this range.
An uncertainty of 2.2 MeV/$c^2$ in the calculated $B_c^{\pm}$ mass is 
indicated due to the fitting procedure.
The combined systematic uncertainty is the sum in quadrature of
the uncertainties in our calibration, tracking, and fitting procedure. 
The total systematic uncertainty in the $B_c^\pm$ mass measurement
is determined to be 2.5~MeV/$c^2$.

In conclusion, we have observed 
fully reconstructed $B_c^{\pm}$ mesons through the decay
$B_c^{\pm} \rightarrow J/\psi \, \pi^{\pm}$.  
A signal of $108\pm15$ candidates is observed with a significance 
greater than an $8\sigma$ fluctuation of a Gaussian distribution.
The mass of the
$B_c^{\pm}$\ meson is measured to be
6275.6~$\pm$~2.9(stat.)~$\pm$~2.5(syst.)~MeV/$c^2$.

We thank the Fermilab staff and the technical staffs of the participating
institutions for their vital contributions. This work was supported by 
the U.S. Department of Energy and National Science Foundation; 
the Italian Istituto Nazionale di Fisica Nucleare; 
the Ministry of Education, Culture, Sports, Science and Technology of Japan;
the Natural Sciences and Engineering Research Council of Canada; 
the National Science Council of the Republic of China; 
the Swiss National Science Foundation; the A.P. Sloan Foundation; 
the Bundesministerium f\"ur Bildung und Forschung, Germany; 
the Korean Science and Engineering Foundation and 
the Korean Research Foundation; 
the Science and Technology Facilities Council and the Royal Society, UK; 
the Institut National de Physique Nucleaire et Physique des Particules/CNRS; 
the Russian Foundation for Basic Research; 
the Comisi\'on Interministerial de Ciencia y Tecnolog\'{\i}a, Spain; 
the European Community's Human Potential Programme; 
the Slovak R\&D Agency; and the Academy of Finland. 

\clearpage

\begin{thebibliography}{99}
\bibitem{pt_def}
The transverse momentum $p_T$ is defined as the component of the particle
momentum perpendicular to the proton  beam direction.
The pseudorapidity $\eta$ is defined as
$\eta \equiv -\ln ( \tan( \theta /2) )$, where $\theta$ 
is the angle between the particle
momentum and the proton beam direction.
\end{thebibliography}
\end{document}